\documentclass{agujournal2019}
\usepackage{url} 
\usepackage{lineno}
\usepackage[inline]{trackchanges} 
\usepackage{soul}
\usepackage{upgreek}
\newcommand{\mum}{$\upmu$m}

\graphicspath{{figures/}}
 \draftfalse

\journalname{JGR: Planets}

\begin{document}

%
%


\title{A Two Martian Years Survey of Water Ice Clouds on Mars with ACS onboard TGO}

%
%




\authors{Aur\'elien Stcherbinine\affil{1, 2}, Franck Montmessin\affil{2}, 
    Mathieu Vincendon\affil{3}, Michael J. Wolff\affil{4}, 
    Margaux Vals\affil{2},
    Oleg Korablev\affil{5}, Anna Fedorova\affil{5}, Alexander Trokhimovskiy\affil{5}, 
    Gaetan Lacombe\affil{2}, Lucio Baggio\affil{2}}

\affiliation{1}{Department of Astronomy and Planetary Science, Northern Arizona University, 
    Flagstaff, AZ 86011, USA}
\affiliation{2}{LATMOS/IPSL, UVSQ Universit\'e Paris-Saclay, Sorbonne Universit\'e, CNRS, 
    78280 Guyancourt, France}
\affiliation{3}{Institut d'Astrophysique Spatiale, Universit\'e Paris-Saclay, CNRS, 91405 Orsay, France}
\affiliation{4}{Space Science Institute, 4750 Walnut Street, Suite 205, Boulder, CO, 80301, USA}
\affiliation{5}{Space Research Institute (IKI), 84/32 Profsoyuznaya, 117997 Moscow, Russia}




\correspondingauthor{Aur\'elien Stcherbinine}{aurelien.stcherbinine@nau.edu}




\begin{keypoints}
\item Cloud altitude varies by 20 to 40~km between winter and summer, and between polar and midlatitudes.
\item The GDS in MY 34 increases the altitude of the clouds by 10 to 20~km compared to the same season in MY 35.
\item Clouds are predicted to be lifted at lower altitudes in the Mars PCM compared to ACS detections.
\end{keypoints}

%
%

%
%


\begin{abstract}
The middle infrared (MIR) channel of the Atmospheric Chemistry Suite (ACS) instrument
onboard the ExoMars Trace Gas Orbiter (TGO) ESA-Roscosmos mission has performed
Solar occultation measurements of the Martian atmosphere in the 2.3--4.2~$\upmu$m
spectral range since March 2018, which now covers two Martian Years (MY).
We use the methodology previously developed for the study of the MY 34
Global Dust Storm (GDS) \cite{stcherbinine_2020a} to monitor the properties (effective radii, extinction, altitude) of the Martian
water ice clouds over the first two Martian years covered by ACS-MIR. The observations encompass the period $L_s=163^\circ$ in MY 34 to $L_s=181^\circ$ in MY 36.
We determine that the typical altitude of the clouds varies by 20 to 40~km between the
summer and winter, with a maximum extension up to 80~km during summer in the midlatitudes.
Similarly, we also note that for a limited temporal range, the altitude of the
clouds also varies by 20 to 40~km between the polar regions and the midlatitudes.
We also compare observations acquired during the MY 34 GDS to observations from the
same period in MY 35, using that latter as a reference to characterize the effects of this GDS
on the clouds' properties.
In addition, we compare our retrievals with the predictions of the Mars Planetary Climate
Model (PCM), which shows a reasonable agreement overall for the altitude of the clouds, although
the model usually predicts lower altitudes for the top of the clouds.
\end{abstract}

\section*{Plain Language Summary}
We use data from the Middle InfraRed (MIR) channel of the Atmospheric Chemistry Suite (ACS)
instrument onboard the ExoMars Trace Gas Orbiter mission, which has been probing the Martian
atmosphere for two Martian years (MY), to study the properties of the water ice clouds in the
Martian atmosphere. We observe that the clouds' altitude increases by 20 to 40~km during Summer and
that they are also observed 20 to 40~km higher around the equator compared to the polar regions.
This highlight the spatial and temporal diversity of the Martian water ice clouds, along with the
large amplitude of the annual variations of the atmospheric structure. In addition, we contrast
observations acquired during the Global Dust Storm (GDS) event of Summer 2018 (MY 34) with data from
the following MY that we use as a reference to understand the effects of the GDS on the clouds,
which reveals that the presence of the GDS increases the altitude of the clouds by 10 to 20~km.
Finally, we compare our results to the predictions of the Mars Planetary Climate Model which shows a
reasonable agreement overall, although the model tends to predict lower altitudes for the top of the
clouds.

%
%

\section{Introduction}  \label{sec:intro}
    Even though the role of water ice clouds in the Martian atmosphere was not well understood for many years 
    after the Viking missions, they are now considered to play a major role
    in Martian climate and weather \cite{richardson_2002, montmessin_2004, 
    navarro_2014a, montmessin_2017, clancy_2017a, vals_2022, rossi_2022}.
    Clouds perform an important role in the Martian water cycle as they are a major actor in the
    inter-hemispheric water exchange \cite{clancy_1996, montmessin_2004, montmessin_2017}. 
    Similarly to atmospheric dust
    particles, water ice crystals also absorb and scatter incoming solar radiation, thus
    impacting the atmospheric structure and temperature \cite{wilson_2007, wilson_2008,
    haberle_2011, madeleine_2012a, navarro_2014a}. In addition, the formation of clouds 
    affects the ability of water (or hydrogen) to be further mobilized and to escape from the planet.
    However, more observational data are needed to better characterize the properties of
    water ice clouds in order to understand and model the evolution of the Martian
    atmosphere \cite{vals_2018a}.
    
    The formation of clouds in the atmosphere depends on several factors, such as the presence of
    water vapor, the pressure and temperature conditions, and the availability of condensation
    nuclei \cite{michelangeli_1993, montmessin_2004}.
    However, as water may exist in a supersaturated state in the Martian atmosphere 
    \cite{maltagliati_2011, fedorova_2020, poncin_2022}, simply considering the freezing point for the
    condensation of atmospheric water is problematic.
    In addition, accurately predicting the distribution and the properties of airborne dust particles
    also remains a challenging aspect in current Global Climate Models (GCM)
    \cite{forget_2014, wang_2018, daversa_2022}.
    Thus, accumulating additional observations of the distribution and properties of the water
    ice crystals in the Martian atmosphere is required to further constrain the present-day clouds
    cycle, and its relationship to the Martian climate and water cycle.
    
    The Atmospheric Chemistry Suite (ACS) instrument is a set of three spectrometers 
    onboard the ExoMars Trace Gas Orbiter (TGO) ESA-Roscosmos spacecraft, which has been conducting
    science operations since March 2018 \cite{korablev_2018, korablev_2019, vandaele_2019}.
    The Mid-InfraRed (MIR) channel is a high-resolution, cross-dispersion
    echelle spectrometer dedicated to solar occultation (hereafter "SO") geometry. Each
    observation covers a $\sim$300~nm wide spectral interval selected between 2.3 and 4.2~\mum,
    which is set by rotating the secondary grating to one of the 10 positions \cite{trokhimovskiy_2015a, korablev_2018}.
    The cross-dispersion optical scheme produces 10 to 21 stacked diffraction orders. 
    The number and separation between the displayed orders depend on the secondary grating
    position.
    Considering the instantaneous angular field of view of the detector and its displacement during the
    integration time for each spectrum, ACS-MIR provides a sampling of the Martian atmosphere with a vertical 
    resolution of $\sim$~2.5~km.
    
    Even though the primary objective of ACS and NOMAD is the study of the atmospheric trace gases
    in the Martian atmosphere, they also provide a unique dataset for the study of the distribution
    and properties of Martian aerosols \cite{stcherbinine_2020a, luginin_2020, liuzzi_2020,
    liuzzi_2021, streeter_2022}.
    In this paper we use ACS-MIR observations acquired in position 12 (i.e., 
    $\sim$3.1--3.4~\mum\ spectral range) to retrieve the properties of the Martian 
    water ice clouds from the 3~\mum\ water ice absorption
    band. Indeed, water ice atmospheric particles exhibit a specific signature due to the O-H
    stretching and bending that enables the distinction between water ice and bound water in dust.
    And as the depth and shape of the absorption feature depend on both the abundance and the sizes of
    the ice crystals, we can retrieve information on these quantities \cite{vincendon_2011,
    guzewich_2014, clancy_2019}.
    The methodology has already been developed and applied to observations conducted during the global dust storm (hereafter "GDS")
    of Martian Year (MY) 34. These efforts revealed very high altitude water ice hazes (up to 100~km) and large ice crystals
    (effective radius $r_\mathrm{eff}\sim$1.5--2~\mum) up to 65~km during the storm \cite{stcherbinine_2020a}.
    ACS observations now cover two MY, one with and one without a GDS, 
    for all latitudes, allowing us to observe the seasonal
    and spatial evolution of the Martian water ice clouds. In addition, observations acquired during
    MY 35 during the same $L_s$ period when the GDS occurs in MY 34 also offer a reference to
    be compared to MY 34 GDS observations.
    
    First, we briefly describe in Section~\ref{sec:data_methods} the ACS-MIR dataset and the methods
    used in this study. Sections~\ref{sec:clouds_monitoring} \& \ref{sec:clouds_kext} present the results of the
    annual monitoring of water ice clouds with their particle sizes and opacity, and
    Section~\ref{sec:comparison_models} compares the retrieved vertical profiles of
    water ice clouds with results of GCM simulations from the
    Mars Planetary Climate Model (PCM) \cite{forget_2022}.
    Finally, Section~\ref{sec:conclusion} summarizes the main points of this study.

\section{Data and methods}  \label{sec:data_methods}
    \subsection{Dataset}    \label{sec:dataset}
        Atmospheric transmittances are computed through ratios to measurements at 120~km above the surface,
        which is free from atmospheric absorption in this spectral range. Indeed, we observed in
        \citeA{stcherbinine_2020a} (where transmittances were computed relative to 150~km)
        that the haze top altitude only reaches 100~km during an extreme event like the MY 34 GDS,
        but without extending over 105~km.
        Data calibration and geometry calculations are described in 
        \citeA{trokhimovskiy_2020} and \citeA{olsen_2021a}.
        
\begin{figure}[h]
            \centering
            \includegraphics[width=\textwidth]{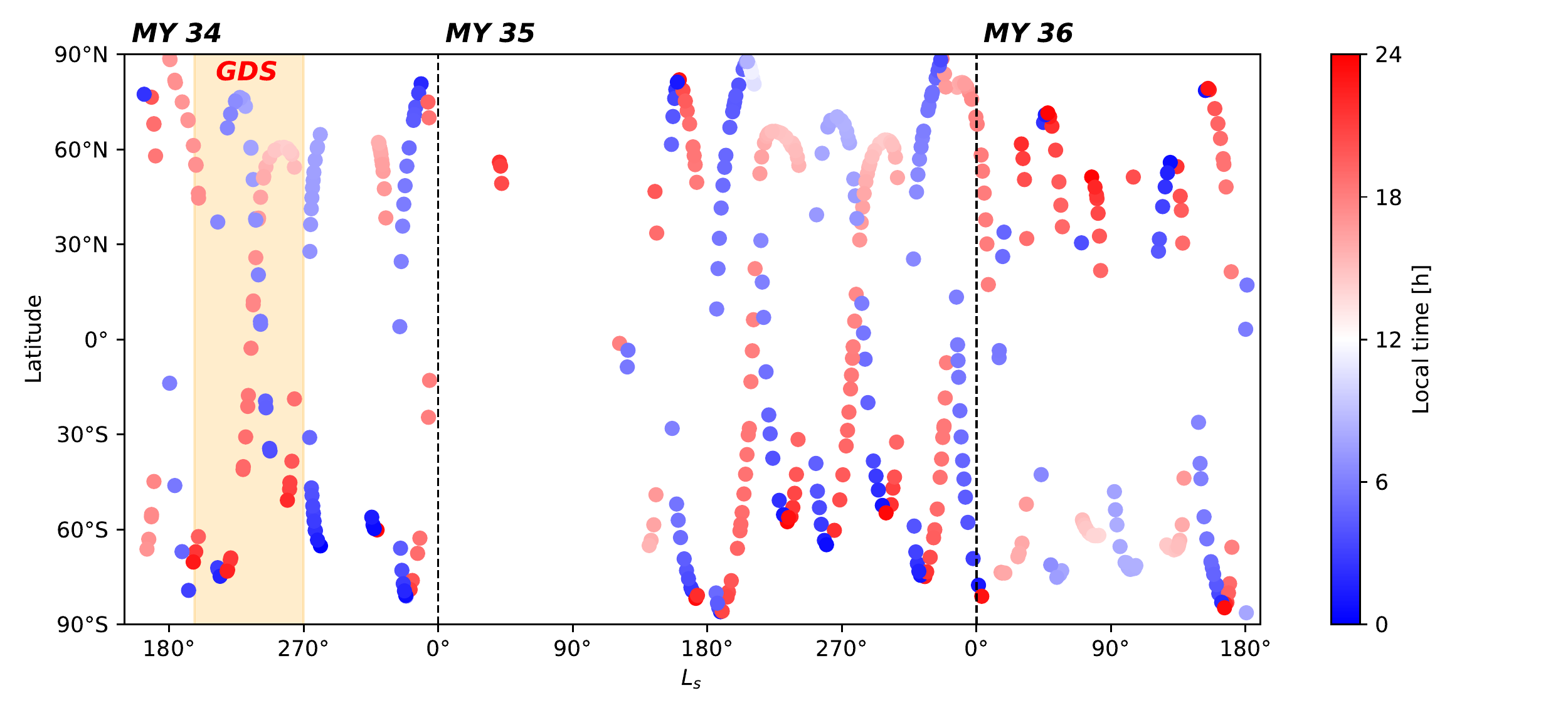}
            \caption{Spatial (latitude) and temporal ($L_s$ and local time) distribution of the
                514 ACS-MIR position 12 observations used in this study.
                The orange region corresponds to the period of the 2018/MY34 global dust storm.}
            \label{fig:distrib_ls_lat_loct_obs_pos12}
        \end{figure}

        Since the publication of \citeA{stcherbinine_2020a}, ACS-MIR has continued to acquire new
        data in the grating position 12, allowing us to observe the atmospheric 3~\mum\ absorption
        band by covering the 3.1--3.4~\mum\ spectral range \cite{korablev_2018}. 
        We do not consider the observations acquired in the so-called
        "partial mode" (i.e., with a smaller number of diffraction orders) around the conjunction of Mars and Earth (when the downlink from the instruments is very limited). A total of 514 observations obtained
        between $L_s=163^\circ$ (MY 34) and $L_s=181^\circ$ (MY 36) were used in this study. 
        The spatial and temporal distribution of these observations is shown in
        Figure~\ref{fig:distrib_ls_lat_loct_obs_pos12}.
        Due to the SO geometry, observations naturally occur in the periods near dawn and twilight.
        
    \subsection{Continuum spectra}    \label{sec:continuum_extraction}
        Based on the methodology previously described in \citeA{stcherbinine_2020a}, we extract the spectral continuum for each
        observed altitude from the 20 spectral segments (i.e.,
        diffraction orders) that compose the ACS-MIR observation \cite{korablev_2018,
        trokhimovskiy_2015a}.
        For each diffraction order, we use successive iterations of a median filter and a second-degree
        polynomial fit on the 200 centered points to determine the transmission level of the spectral
        continuum at the center of the diffraction order, while minimizing the effects of contamination by the gas absorption
        bands or the broad instrumental curvature that affects the orders.
        This results (for each observed altitude) in a spectrum with one point per
        diffraction order, with a spectral resolution of $\Delta \lambda \sim 20~\mathrm{nm}$. To avoid
        detector edges effects, we only consider the center of each diffraction order to process the
        extraction. We also ignore the outer regions of the extracted spectra, that is, the
        spectels corresponding to 3.10, 3.12, and 3.44~$\upmu$m.
        
        We have updated the uncertainties estimation for the ACS-MIR transmittances. Thus, as we are
        considering the center of each diffraction order, we include an uncertainty of $\pm 5\%$ of the
        transmission value to account for the uncorrected straylight that affects the data in position
        12 \cite{trokhimovskiy_2020}.
        
        We compute the haze top altitude (i.e., the highest altitude at which aerosols can be detected
        along the line of sight), defined here as the first altitude for which the transmission is
        greater than $0.98$ over the entire spectrum, along with the extinction coefficient
        $k_\mathrm{ext}$ for each wavelength (forming an extinction spectrum) 
        and altitude from the total optical depth (i.e., integrated along the line of sight, 
        $\tau = -\log (\mathrm{Tr})$) using a vertical profile inversion algorithm based on the
        onion-peeling method \cite{goldman_1979} as described in sections 2.2 and 2.4 of \citeA{stcherbinine_2020a}.
        
    \subsection{Water ice particle size retrieval method}
    \label{sec:retrieving_method}
        To identify the Martian water ice clouds in our observations, and constrain their particle
        sizes, we compare each retrieved extinction spectrum with models of the extinction
        coefficient's wavelength dependence for either water ice or dust spherical particles
        \cite{stcherbinine_2020a}.
        These theoretical extinction coefficients are computed using a public domain Mie code
        \cite{toon_1981} and assuming a gamma size distribution \cite{hansen_1974} with an
        effective variance of 0.1 \cite<e.g.,>[and references contained within]{wolff_2017}.

        To avoid the problem of the presence of local minima for $r_\mathrm{eff}$ during the fitting
        process, we generate models on a grid of radii of 0.01 between 0.1 and 8~\mum, and for each one we
        compute the $\chi^2$ and $\Delta^2$ values defined in equations~(\ref{eq:chi2nu}) \& (\ref{eq:delta2}), which quantify the deviation between the data and the model.
        Finally, we have slightly adjusted the criterion 
        for the detection and characterization of the atmospheric water ice particles compared to \citeA{stcherbinine_2020a}. 
        This modification results from a better understanding of detection biases thanks to the new larger dataset, 
        and also accounts for the new uncertainty estimate (section~\ref{sec:continuum_extraction}). The new criterion is as follows:
        
        \begin{linenomath*}
        \begin{equation}    \label{eq:critere}
            \left( \chi^2_{\nu,\, \mathrm{ice}} < 1 \right) \&
            \left( \frac{\Delta^2_\mathrm{dust}}{\Delta^2_\mathrm{ice}} > 4 \right)
        \end{equation}
        \end{linenomath*}
        
        Where
        \begin{linenomath*}
        \begin{equation}    \label{eq:chi2nu}
            \chi^2_\nu(r_\mathrm{eff}) = \frac{1}{N - 2}
                    \sum_{i=1}^N \frac{\left(\textrm{data}_i - \textrm{model}_{r_\mathrm{eff},\,i}\right)^2}
                                            {\sigma_i^2}
        \end{equation}
        \end{linenomath*}
        
        and
        \begin{linenomath*}
        \begin{equation}    \label{eq:delta2}
            \Delta^2(r_\mathrm{eff}) = \frac{1}{N - 2}
                    \sum_{i=1}^N \left(\textrm{data}_i - \textrm{model}_{r_\mathrm{eff},\,i}\right)^2
        \end{equation}
        \end{linenomath*}
        
        \begin{notation}
            \notation{$\mathrm{data}_i$} The $i^\mathrm{th}$ spectel of the $k_\mathrm{ext}$ spectra
                from the ACS-MIR observation.
            \notation{$\mathrm{model}_{r_\mathrm{eff},\,i}$} The $i^\mathrm{th}$ spectel of the model 
                extinction spectra for a particle size of $r_\mathrm{eff}$.
            \notation{$\sigma_i$} The uncertainty on the value of data$_i$.
            \notation{$N$} The number of spectral points in the considered spectrum.
        \end{notation}
        
        Specifically, we require that:
        \begin{enumerate}
            \item the modeled extinction spectrum using water ice provides a good fit to the observational
                data, including the spectrum uncertainties 
                $\left( \chi^2_{\nu,\, \mathrm{ice}} < 1 \right)$.
            \item the water ice model provides a significantly better solution than the best fit that 
                can be obtained with a dust model. So we consider only water ice models with a mean square
                difference to the data lower than the one of the best dust model by at least a factor 4.
                This factor 4 has been determined experimentally by visual comparison between the
                models and the data for many spectra, to retain only those which match within the
                errorbars.
        \end{enumerate}
        
        For each spectrum where water ice crystals are detected, we
        retrieve the optimal $r_\mathrm{eff}$ as the one corresponding to the model associated with
        the lower $\chi^2_{\nu,\, \mathrm{ice}}$ that verifies equation~(\ref{eq:critere}).
        Then, the lower (respectively upper) bound for the particle size uncertainties corresponds to
        the minimum (respectively maximum) value of $r_\mathrm{eff}$ in the set of models 
        associated with $\chi^2_{\nu,\, \mathrm{ice}}$ and $\Delta^2_\mathrm{ice}$ that prove
        equation~(\ref{eq:critere}).

\section{Cloud monitoring}     \label{sec:clouds_monitoring}
        In sections~\ref{sec:seasonal_variations} and \ref{sec:latitudinal_variations}, we present the results of the analysis of the new data, as described in \citeA{stcherbinine_2020a}.
        They cover a complete year without a GDS, from $L_s=140^\circ$
        (MY 35) to $L_s=182^\circ$ (MY 36). This range corresponds to ACS-MIR operating in full-observation mode (cf.
        Figure~\ref{fig:distrib_ls_lat_loct_obs_pos12}). Then, we compare this new dataset to the previous one for the GDS year MY 34 in section~\ref{sec:comp_GDS34}.

        In the following, a "profile" refers to one set of ACS-MIR observations from the surface to the
        haze top, while a "cloud" corresponds to contiguous water ice detections within a profile.
        Additionally, as we observe the presence of water ice clouds up to the haze top altitude
        in most of the profiles, we consider the haze top equivalent to the maximum altitude of the water
        ice clouds. Also, our water ice clouds detections extend typically over $\sim$~20~km below their
        maximum altitude, thus we will only discuss in this section the maximum altitude of the water
        ice clouds as it is representative of the overall behavior of the water ice clouds (and the
        maximum altitude of the aerosols in general).

    \subsection{Seasonal variations}    \label{sec:seasonal_variations}
\begin{figure}[htp]
            \centering
            \includegraphics[width=\textwidth]{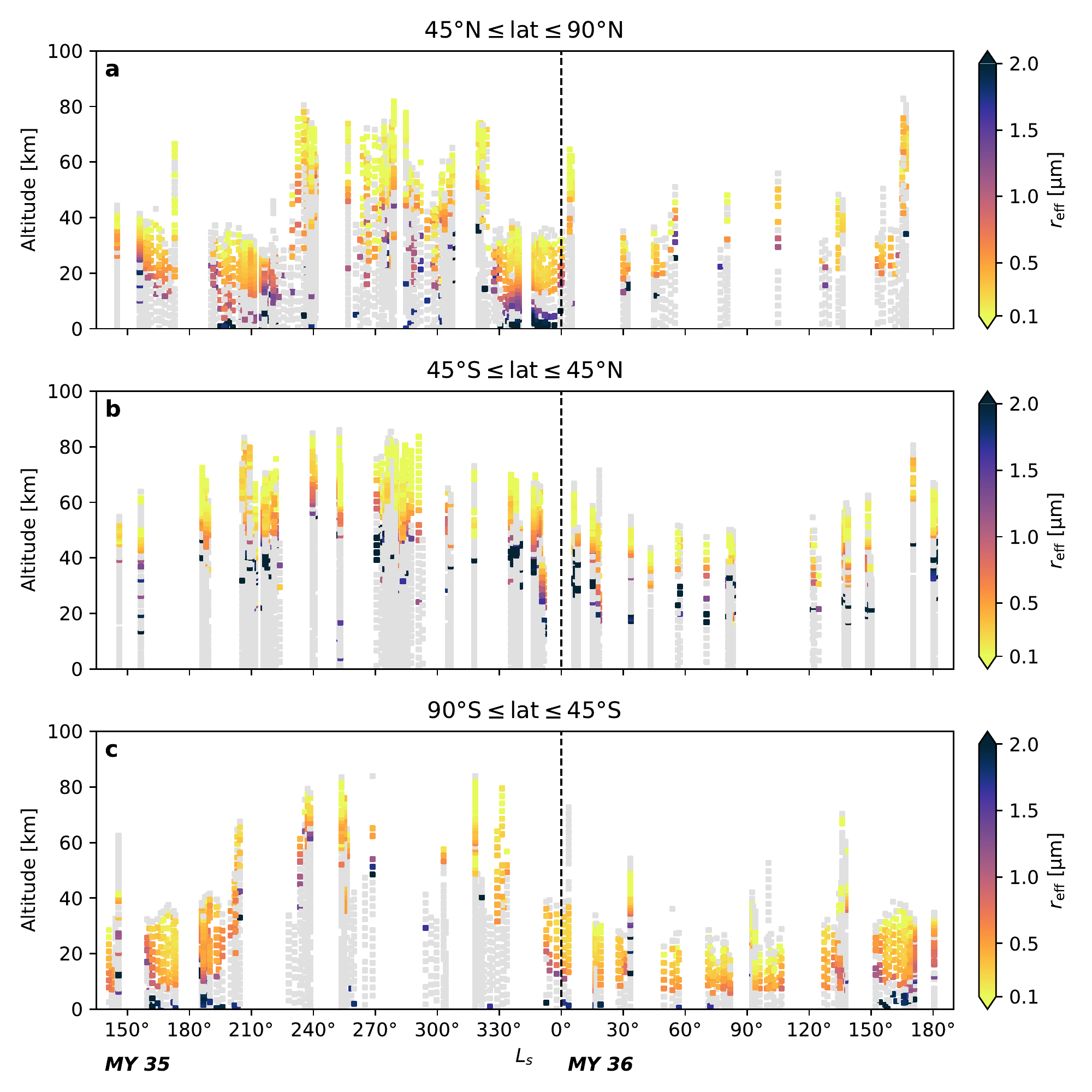}
            \caption{Vertical profiles of water ice clouds in the Martian atmosphere as observed by
              ACS-MIR over mid-MY 35 to mid-MY 36, with their crystal size determined
              using the method presented in section~\ref{sec:retrieving_method}.
              Each panel represents the profiles grouped by ranges of latitude
              (North/Equatorial/South). Observations without water ice detections are in gray.}
            \label{fig:profils_reff_my35-36_filtrage_lat}
        \end{figure}

        We show seasonal variations for three distinct latitude ranges in
        Figure~\ref{fig:profils_reff_my35-36_filtrage_lat} in order to delineate the seasonal variations 
        from the latitudinal dependency of the clouds that will be discussed in paragraph~\ref{sec:latitudinal_variations}. 
  
        As reported in previous studies \cite<e.g.,>{clancy_2019,
        stcherbinine_2020a, luginin_2020, liuzzi_2020}, we find a decrease in the size of the
        water ice crystals as altitude increases, both from an individual cloud and from a global perspective
        (cf. Figures~\ref{fig:profils_reff_my35-36_filtrage_lat},
        \ref{fig:profils_reff_my35-36_filtrage_ls} \& \ref{fig:alt_reff_my35-36_filtrage_ls}). 
        In addition, clouds are usually found at the top of the profiles, capping the layers composed of other
        types of aerosols (dust or large ice crystals below).

        We also observe in Figure~\ref{fig:profils_reff_my35-36_filtrage_lat} that if the maximum altitude of the clouds fluctuates with season \cite{jaquin_1986,
        forget_1999, montmessin_2006, heavens_2011a, maattanen_2013, smith_2013}, the observed
        variations depend on the latitude.
        Indeed, at equatorial latitudes ($45^\circ$S $\leq$ lat $\leq$ $45^\circ$N, panel b), we observe
        the presence of water ice crystals up to 83~km around the perihelion ($L_s\sim270^\circ$),
        while the maximum altitude of the clouds does not exceed 50~km at $L_s\sim80^\circ$. Plus, for latitudes
        southward $45^\circ$S (panel c) we note an increase of the maximum altitude of the clouds
        that goes from 30~km at $L_s\sim160^\circ$ to 80~km at $L_s\sim260^\circ$,
        followed by a decrease from $L_s\sim330^\circ$ that brings it back to 30~km at
        $L_s\sim20^\circ$.
        Symmetrically, in the Northern hemisphere, between $45^\circ$N and $90^\circ$N (panel a)
        the maximum altitude of the clouds moves from 35
        to 20~km between $L_s=145^\circ$ and $L_s=215^\circ$, and from 20 to 40~km between
        $L_s=30^\circ$ and $L_s=100^\circ$ before coming back to 25~km at $L_s=160^\circ$.
    
        \subsubsection*{Observation of a regional dust storm}    \label{sec:regional_DS}
            In addition to these variations of the water ice clouds
            altitude with $L_s$, we observe in the Northern hemisphere (for latitudes greater
            than $45^\circ$N; Figure~\ref{fig:profils_reff_my35-36_filtrage_lat}a) a sudden increase
            of their maximum altitude between $L_s=236^\circ$ and $L_s=323^\circ$, which indicates
            the presence of a regional dust storm in this region.
            This dust storm and its impact on the altitude of the water ice clouds have also been
            observed by the NOMAD instrument and are reported in \citeA{streeter_2022}.
            Indeed, from
            $L_s=225^\circ$ to $L_s=240^\circ$ the maximum altitude of the clouds goes from 35 to 80~km.
            That is to say, there is an increase in the altitude of the water ice clouds of about
            40~km while the latitude of the observations remains
            similar (between $60^\circ$N and $65^\circ$N, cf Figure~\ref{fig:distrib_ls_lat_loct_obs_pos12}).
            The clouds remain in these high atmospheric layers (with crystals observed up to 80~km)
            until $L_s=325^\circ$, even though we can observe a progressive decrease of the maximum
            altitude of the clouds over the storm, which goes from 80 to 65~km between
            $L_s=235^\circ$ and $L_s=305^\circ$ while the observed latitudes oscillate between
            $41^\circ$N and $63^\circ$N.
            Then, the maximum altitude of the water ice clouds suddenly decrease between $L_s=323^\circ$ and
            $L_s=328^\circ$, going from 71 to 28~km. However, as $10^\circ$ of latitude separate
            these two observations ($61.5^\circ$N for $L_s=323^\circ$ and $71.5^\circ$N for
            $L_s=328^\circ$, 
            plus, $71.5^\circ$N is above the northernmost latitude of the observations of the storm),
            it is challenging to determine precisely the end date of the dust storm with our observations. 
            For example, the northern limit of the MY 34 GDS was very abrupt and located under similar latitudes \cite{stcherbinine_2020a}.
            In addition, $71.5^\circ$N is also close to the typical boundary of the north polar vortex at this period \cite{ball_2021}, which may act as an efficient barrier to the latitudinal transport of atmospheric particles \cite{toigo_2020}.

            Regarding the temporal behavior of this dust storm, we observe that it occurs in the second half of the "dust storm season", which illustrates the diversity of Martian dust storms
            and the necessity to observe and study a large number of these events \cite{wang_2015, battalio_2021}.
            We also observe that this storm does not push clouds to altitudes as high as what has
            been observed during the MY 34 GDS (80~km here vs 100~km during the GDS), and that no
            water ice crystals larger than 1.5~\mum\ are observed above 40~km, unlike during the GDS
            where such large crystals have been detected between 50 and 65~km.

    \subsection{Latitudinal variations} \label{sec:latitudinal_variations}
        \begin{figure}[ht]
            \centering
            \includegraphics[width=\textwidth]{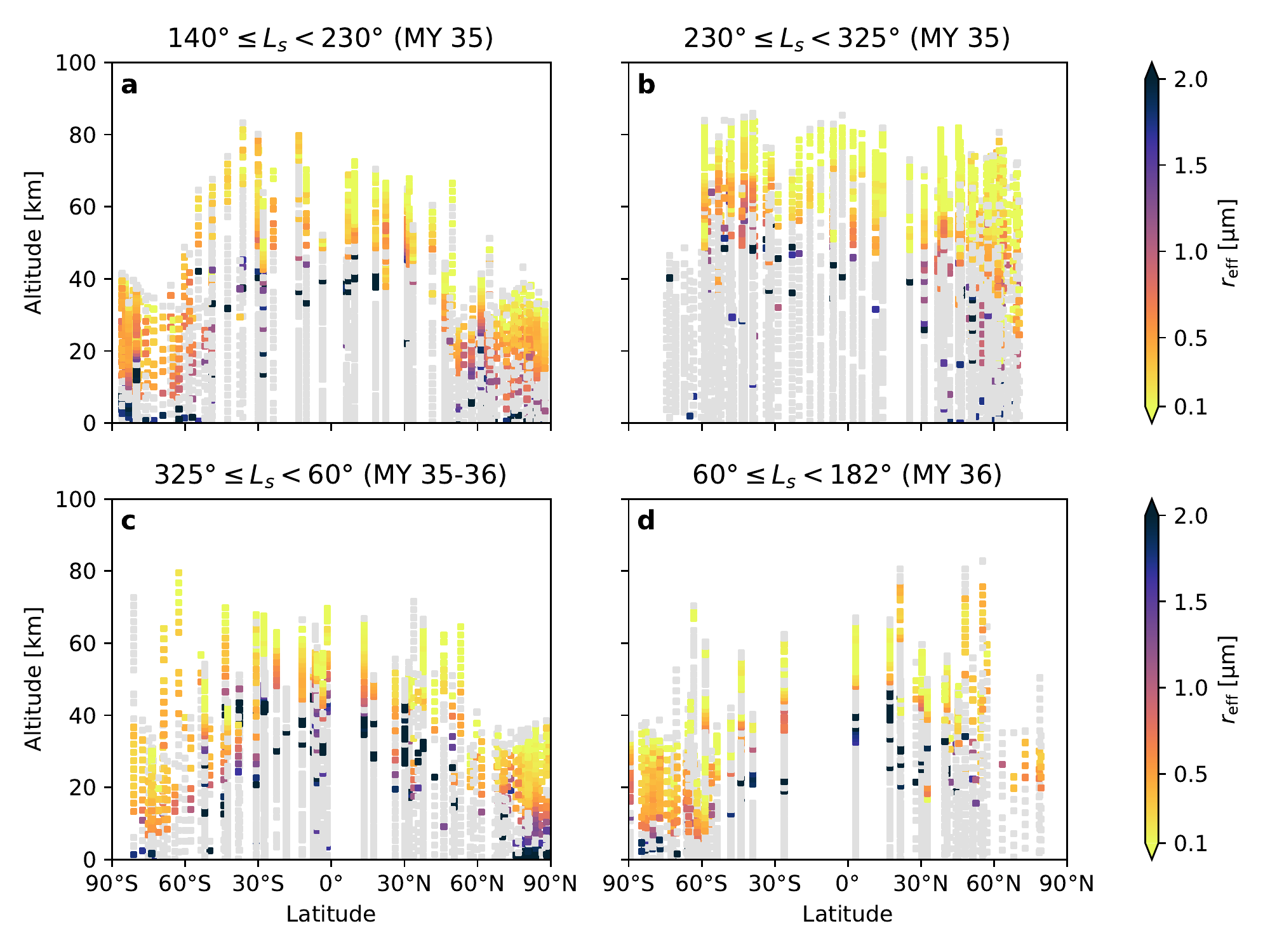}
            \caption{Vertical profiles of water ice clouds in the Martian atmosphere as observed by
              ACS-MIR between $L_s=140^\circ$ (MY 35) and $L_s=182^\circ$ (MY 36) as a function of
              the latitude of the observation. To highlight the latitudinal variations of
              the clouds while removing their seasonal dependence (discussed in
              section~\ref{sec:latitudinal_variations}), each panel represents the profiles grouped
              by $L_s$ ranges.
              Observations without water ice detection are in gray.}
            \label{fig:profils_reff_my35-36_filtrage_ls}
        \end{figure}
    
        \begin{figure}[htp]
            \centering
            \includegraphics[width=\textwidth]{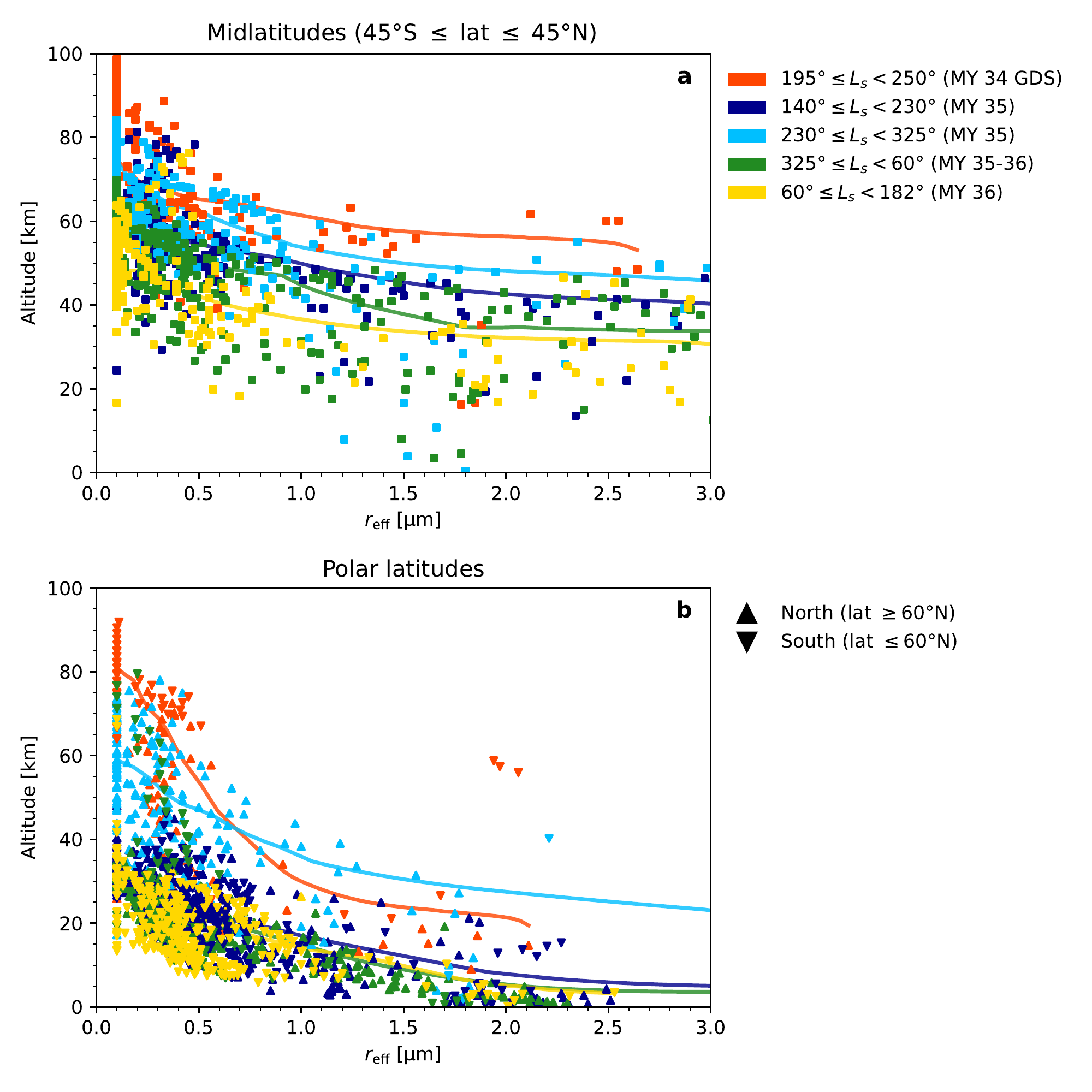}
            \caption{Distribution and size of the water ice crystals within the clouds as a function
              of their altitude and $L_s$ of observation, for 3 ranges of latitude: midlatidudes
              (panel a), Northern and Southern polar latitudes (panel b).
              The solid lines show the average trend for each range of $L_s$
              Observations acquired during the MY 34 GDS are shown in red for comparison with data
              from non-GDS years (MY 35 \& 36).}
            \label{fig:alt_reff_my35-36_filtrage_ls}
        \end{figure}
    
        As mentioned above, and reported in previous studies \cite{jaquin_1986,
        forget_1999, montmessin_2006, heavens_2011a, smith_2013} the vertical structure of Martian
        aerosols depends on the latitude.
        To isolate this latitudinal dependence from the seasonal variation discussed in
        section~\ref{sec:seasonal_variations}, Figure~\ref{fig:profils_reff_my35-36_filtrage_ls}
        presents the vertical profiles of  water ice crystals size within the clouds. We adjust several $L_s$ bins to minimize the impact of the seasonal
        variations while providing enough coverage in terms of latitude.
        If the entire range of latitudes is not covered in each panel, we can observe on
        panels a, c \& d that the altitude of the water ice clouds (and more generally of the
        Martian aerosols) increases when getting closer to the equator. This also produces a
        bell-shaped distribution of the altitude of the clouds as a function of latitude. Indeed, we
        observe on Figures~\ref{fig:profils_reff_my35-36_filtrage_ls} \&
        \ref{fig:alt_reff_my35-36_filtrage_ls} that if the water ice crystals with radius
        $r_\mathrm{eff}\leq1$~\mum\ are located between 10 and 40~km in the polar regions
        ($>60^\circ$N or $<60^\circ$S), they are found typically between 30 and 80~km around the
        equator, and occasionally up to 85~km.
  
        Another noteworthy point regarding the size of the water ice crystals that compose the
        clouds detected by ACS-MIR: we observe that for a fixed size of particle, the altitude of the 
        detections still follows the
        bell-shaped distribution with latitude already identified for the general distribution of the clouds in the previous paragraph.
        Thus, there is no strict correlation between the size of a crystal and its altitude in the
        atmosphere. Other parameters have to be considered, such as the latitude of the cloud. This
        is consistent with the decrease of the scale height of the atmosphere in the high latitudes,
        where the temperatures are lower. 
        As the temperatures decrease with increasing latitude, one finds a decrease in the cloud altitudes.
        However, we observe in Figure~\ref{fig:alt_reff_my35-36_filtrage_ls} that we also do not
        detect crystals with radius $\geq1.5$~\mum\ above 55~km (outside the MY 34 GDS), there is no
        crystal with size $\leq0.2$~\mum\ at altitudes lower than 15~km. Plus, except for one
        atypical cloud observed by $35^\circ$S of latitude and $252^\circ$ of $L_s$ (MY 35), only
        crystals with sizes $\leq0.1$~\mum\ populate the altitudes higher than 80~km. Thus, we can
        identify with Figure~\ref{fig:alt_reff_my35-36_filtrage_ls} the range of altitudes that are
        populated by water ice crystals for a given radius.
  
        Regarding the larger crystals ($r_\mathrm{eff}\geq1.5$~\mum), they dominate the
        lower layers of the clouds and are observed up to 55~km of altitude around $30^\circ$S.
        However, such particles are not observed for all profiles in which a water ice cloud has
        been detected, unlike small-grained high altitude hazes that are observed at the top of
        almost every profile.
        The localized aspect together with the observed altitudes of these clouds composed
        of crystals with $r_\mathrm{eff}\sim2$~\mum\ is consistent with the results obtained
        using CRISM limb observations presented in figures 10 to 17 of \citeA{smith_2013}.
  
        Concerning the local dust storm occurring in the Northern hemisphere during MY 35 (cf.
        section~\ref{sec:regional_DS}), panel b of Figure~\ref{fig:profils_reff_my35-36_filtrage_ls}
        presents profiles of clouds acquired during this period. This event allows us to study its specific impact
        on the clouds in the Northern latitudes in relation to the rest of the planet.
        Although TGO's orbit over this period did not allow ACS to probe latitudes above
        $65^\circ$S and $70^\circ$N, we observe that the maximum altitude of the clouds observed
        near $70^\circ$N extends to 70~km, which is higher than clouds observed at these
        latitudes outside of the storm (cf. panels a, c \& d of
        Figure~\ref{fig:profils_reff_my35-36_filtrage_ls}), but lower than the altitudes of
        the mesospheric clouds between $40^\circ$S and $20^\circ$N \cite{clancy_2019}.
    
    \subsection{Comparison with the 2018/MY34 GDS}  \label{sec:comp_GDS34}
        \begin{figure}[htbp]
            \centering
            \includegraphics[width=\textwidth]{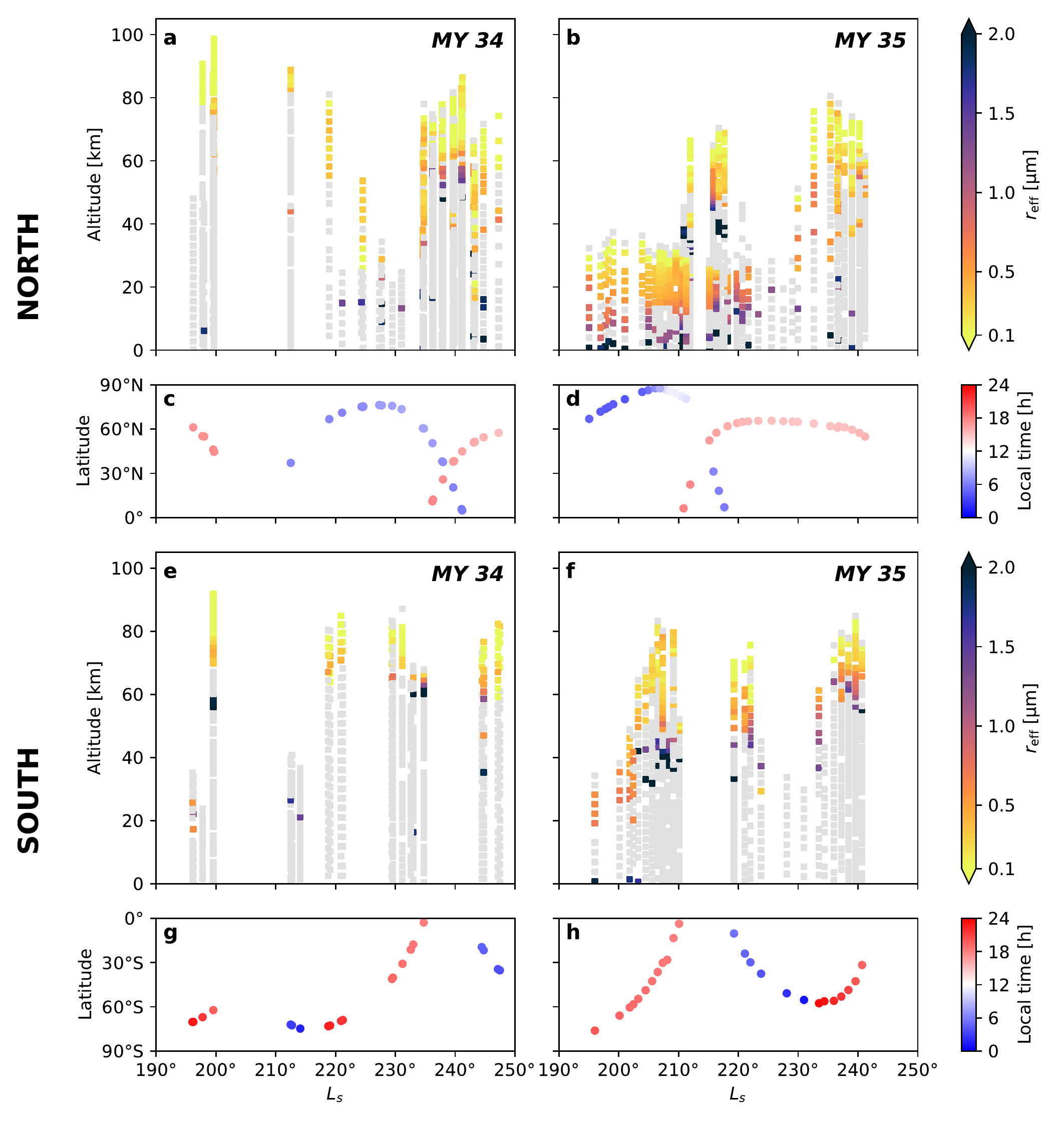}
            \caption{Vertical profiles of water ice clouds in the Martian atmosphere as observed by
              ACS-MIR between $L_s=190^\circ$ and $L_s=250^\circ$ over MY 34 (a \& e) and MY 35 (b
              \& f). Panels (c, d, g \& h) indicate the latitudes and local times of the
              observations presented in panels (a, b, e \& f) respectively.
              Observations without water ice detection are in gray.}
            \label{fig:profils_reff_comp_GDS_my34-35}
        \end{figure}
    
        \citeA{stcherbinine_2020a} used the early observations provided by ACS-MIR to
        study the planetary-scale dust storm that encircled Mars during MY 34.
        With the current extended dataset, it is now possible to compare observations acquired during the MY
        34 GDS with the ones obtained for the same range of $L_s$ in MY 35, when there was no GDS.
  
        Figure~\ref{fig:profils_reff_comp_GDS_my34-35} presents a superimposition of the profiles
        obtained between $L_s=190^\circ$ and $L_s=250^\circ$ during MY 34 (panels a \& e) and MY 35
        (panels b \& f). The latitudes and local times of the observations are indicated in panels
        c, d, g \& h respectively. This information should be taken into account when comparing
        the two years, as the latitude of the clouds has a significant impact on its
        profile characteristics (cf. section~\ref{sec:latitudinal_variations}).
        Even when the two datasets cover the same latitude range, we note in
        Figures~\ref{fig:alt_reff_my35-36_filtrage_ls} \& \ref{fig:profils_reff_comp_GDS_my34-35} that
        the altitude of the clouds for MY 35 does not exceed 85~km (consistent with the results
        presented in \citeA{clancy_2019}), while small-particle clouds ($r_\mathrm{eff}\leq0.1$~\mum)
        are observed up to 100~km at the beginning of the GDS. 
        In particular, we observe at the onset of the GDS ($L_s=195.5^\circ$, MY 34)
        a profile in the Southern hemisphere ($62^\circ$S) with a cloud that extends up to 92~km along with large
        water ice crystals ($r_\mathrm{eff}\sim2$~\mum) around 58~km, which contrasts with the
        observations acquired under the same conditions during MY 35 with clouds that do not
        extend over 40 to 50~km.
         
        The period around $L_s\sim240^\circ$ in the Northern hemisphere is also interesting, It
        not only corresponds to the end of the MY 34 GDS, but also to the maximum of dust activity
        for a regular (i.e., without GDS) year \cite{lemmon_2015}. We also note the presence of a
        regional dust storm during this period in MY 35 (cf. section~\ref{sec:regional_DS}).
        We observe that the maximum altitudes of the clouds are quite similar in both cases, i.e., between
        70 and 85~km. However, we do not observe the large water ice crystals
        ($r_\mathrm{eff}>1$~\mum) during the MY 35 dust storm that were seen between 50 and 60~km in MY 34 during the
        GDS.
  
        Regarding the larger water ice crystals ($r_\mathrm{eff}\sim1.5-2$~\mum), whose detection
        between 55 and 65~km during the GDS was surprising \cite{stcherbinine_2020a}, similar sizes
        can be observed up to 55~km in the equatorial regions in MY 35 (i.e., no GDS).
        However, these larger crystals generally remain confined to altitudes below 50~km.
        For smaller particle sizes ($r_\mathrm{eff}\sim1-1.5$~\mum), we observe that while the detections
        during the GDS are mostly concentrated around 60~km of altitude, they are typically found
        near 50~km during the same period in MY 35 (cf. Figure~\ref{fig:alt_reff_my35-36_filtrage_ls}).
        Similarly, we also observe in Figure~\ref{fig:profils_reff_comp_GDS_my34-35} that the detections 
        of water ice crystals with sizes 0.13~\mum\ $\leq r_\mathrm{eff} \leq$ 0.5~\mum\
        that are detected between 75 and 85~km during the GDS (panel a) are not present in the following
        year (panel b), where this size of crystals is not observed above 75~km.
        Thus, the increase of the average altitude of the clouds during the GDS does not affect all
        the ice particles in the same way, rather it depends on their size: the increase of the maximum
        altitude of the smallest crystals ($r_\mathrm{eff}\leq0.1$~\mum) is
        20~km more during the storm, and only 10~km for those between 0.1 and 0.5~\mum.
  
        We also observe that the impact of the MY 34 GDS on the clouds and aerosols is
        stronger at high latitudes than in the equatorial region. Indeed, it has been
        reported that the altitude of clouds does not vary as a function of latitude during the
        GDS, which tends to homogenize the vertical structure of the atmosphere across the planet
        \cite{stcherbinine_2020a, luginin_2020, liuzzi_2020}. This effect is particularly visible in
        the Southern hemisphere (panels e to h of Figure~\ref{fig:profils_reff_comp_GDS_my34-35}).
        Thus, while the typical maximum altitude of equatorial clouds varies from $\sim80$~km to
        $\sim90$~km during the GDS, it moves from $\sim40$ to $\sim80$~km for latitudes
        poleward of $60^\circ$S and $60^\circ$N.
        In addition, Figure~\ref{fig:alt_reff_my35-36_filtrage_ls}b shows that the detections of 2~\mum\
        ice crystals at 60~km during the GDS in the Southern latitudes are significantly higher than all
        the other detections in MY 35 that usually do not exceed 20~km of altitude for this particle size.
        Although one exception is a cloud with crystals of $r_\mathrm{eff}=2.2$~\mum\
        detected at 40~km in the Southern hemisphere at $L_s=321^\circ$ (MY 35), which is still 20~km
        below the altitude where such clouds were detected during the GDS.
    
\section{Aerosols extinction}  \label{sec:clouds_kext}
    \begin{figure}[ht]
        \centering
        \includegraphics[width=\textwidth]{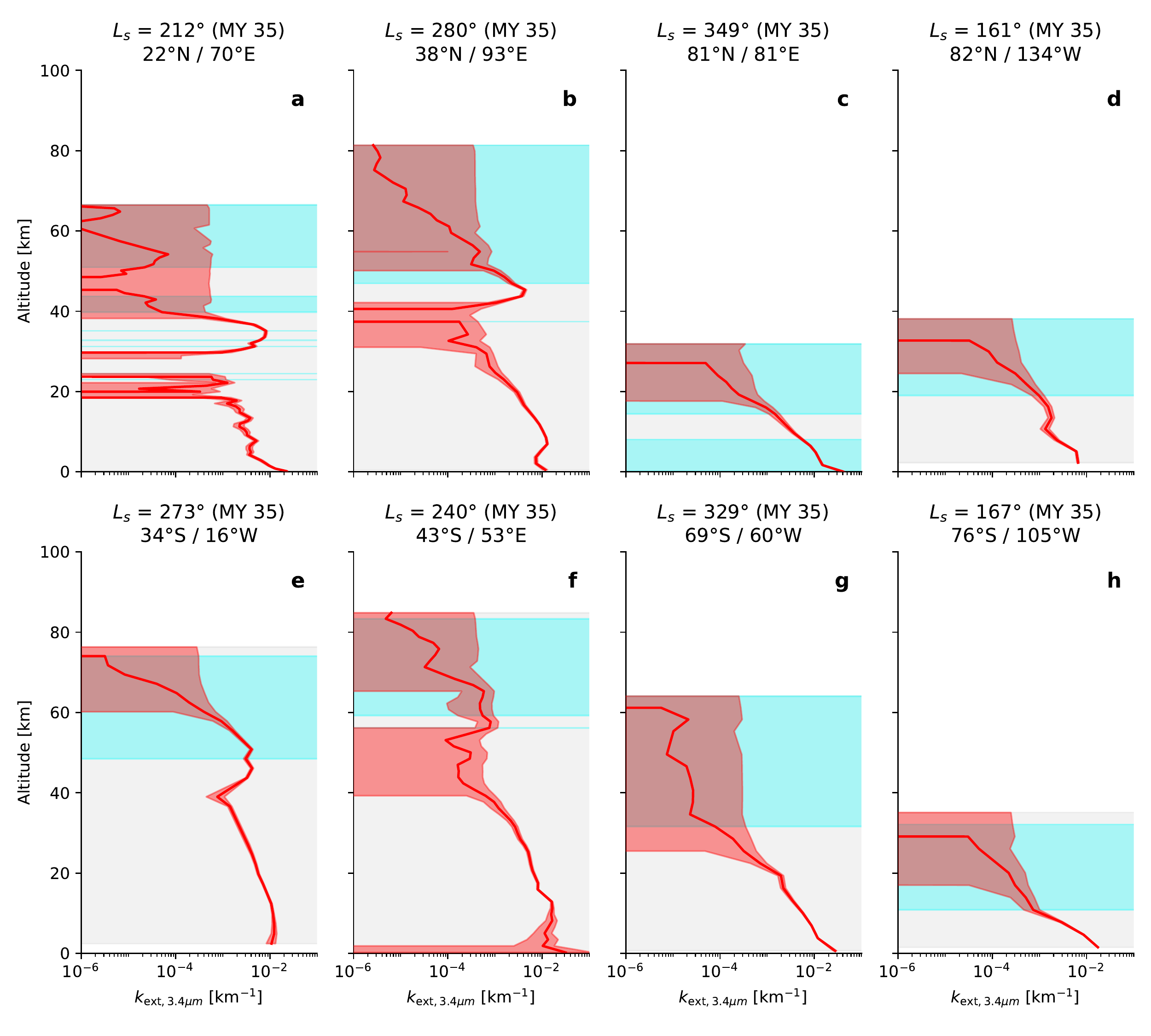}
        \caption{Vertical profiles of the derived extinction coefficient $k_\mathrm{ext}$ 
          at $\lambda=3.4$~\mum\ from 8 different ACS-MIR observations acquired during MY 35 at
          various latitudes.
          The light blue areas represent the altitudes where water ice clouds have been 
          identified.}
        \label{fig:multi_kext_profiles}
    \end{figure}
        
    \subsection{Study of individual profiles}
        Along with particle size, the extinction coefficient is an important property. It will influence the amount
        of incoming solar energy that will be thermally transferred to the atmosphere 
        \cite{gierasch_1972, montmessin_2002a, madeleine_2011, madeleine_2012a}.
        The SO geometry used by ACS-MIR gives us direct access to the vertical profiles of the
        atmospheric extinction at each observed wavelength through the vertical inversion
        technique. In fact, as described in Section~\ref{sec:data_methods}
        and \citeA{stcherbinine_2020a}, this step has already been performed as part of our derivation of the
        properties of the water ice clouds.

        Figure~\ref{fig:multi_kext_profiles} shows 8 extinction profiles at $\lambda=3.4$~\mum\
        acquired by ACS-MIR at various latitudes during MY 35, showing altitudes where water
        ice particles have been detected in each profile. 
        Because 3.4~\mum\ is located on the edge of the 3~\mum\ band, it is less sensitive to 
        the composition of the atmospheric layer (water ice or dust) and to the size of the water
        ice crystals \cite{vincendon_2011}. As a result, it is more indicative of the general aerosol extinction of the
        atmosphere (compared to the extinction at
        3.2~\mum\ that highlight the presence of small-grained water ice crystals).  
        We observe a difference between polar latitudes (c, d, g \& h) and
        more tropical or equatorial observations (a, b, e \& f).
        Indeed, while the main cloud layers are typically observed above 50~km in the equatorial
        profiles, we also note the presence of additional layers at lower altitudes (around 10~km).
        This is particularly clear in profile~b, where the extinction goes from 
        $7\cdot10^{-3}$~km$^{-1}$ at 3~km to $1.3\cdot10^{-2}$~km$^{-1}$ at 7~km before progressively
        decreasing to $10^{-4}$~km$^{-1}$ at 40~km.
        Then, the extinction increases again up to $4\cdot10^{-3}$~km$^{-1}$ between 40~km and 52~km,
        which corresponds to the bottom of the water ice layer that extends from 47~km to 82~km.
        The presence of multiple high-extinction layers has
        been previously observed by SPICAM (Spectroscopy for the Investigation of the
        Characteristics of the Atmosphere of Mars) onboard Mars Express 
        \cite{fedorova_2009, fedorova_2014}.
        
        We also note in profiles b \& e that we only detect water ice crystals in the upper part
        of the detached layers and above, which means that the lower altitudes of these layers are
        primarily composed of either dust or large water ice crystals.
        Similarly, we observe in profile~a some detections of large ($r_\mathrm{eff}=1.5-3$~\mum)
        water ice crystals between 23~km and 36~km of altitude, which suggests the presence of
        water ice and not only dust in these lower layers.
        This vertical structure of ice layers capping the dust layers has already been reported 
        with limb observations \cite{smith_2013}.
        In addition, clouds are detected at the top of the profiles associated with extinction down
        to a few $10^{-5}$~km$^{-1}$. The detection of these very tenuous water ice hazes is made
        possible by the high sensitivity provided by ACS-MIR and the SO geometry technique.
        
    \begin{figure}[ht]
        \centering
        \includegraphics[width=\textwidth]{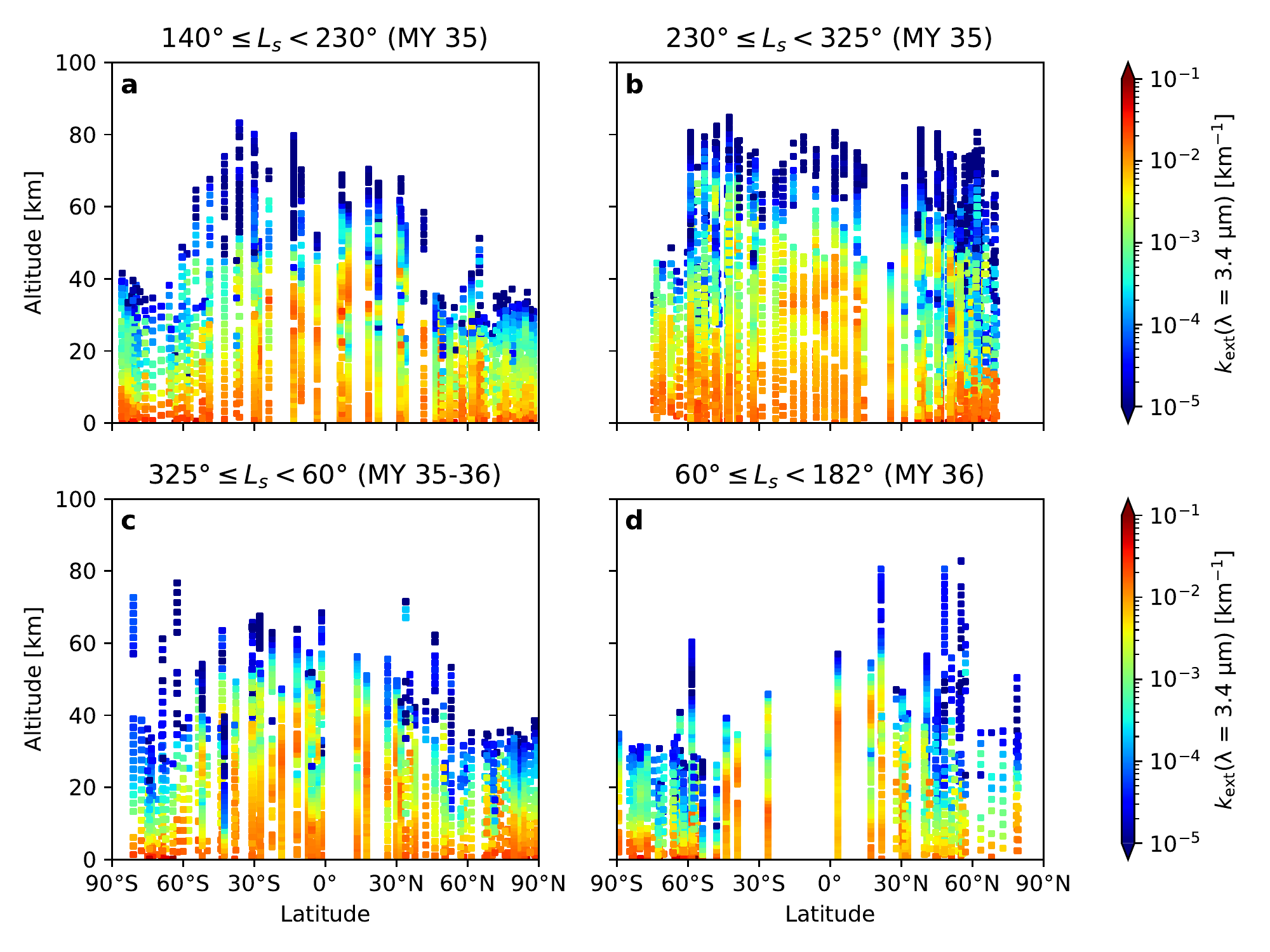}
        \caption{Vertical profiles of the measured extinction of the Martian atmosphere
          ($k_\mathrm{ext}=\mathrm{d}\tau/\mathrm{d}z$, in km$^{-1}$) at 3.4~\mum\ as observed
          by ACS-MIR between $L_s=140^\circ$ (MY 35) and $L_s=182^\circ$ (MY 36) as a function of
          the latitude of the observation. To highlight the latitudinal variations of the clouds
          by getting rid of their seasonal dependency (discussed in
          section~\ref{sec:latitudinal_variations}), each panel represents the profiles grouped
          by ranges of $L_s$.}
        \label{fig:profils_kext_my35-36_filtrage_ls}
    \end{figure}
        
    \subsection{Latitudinal variations}
\begin{figure}[htp]
        \centering
        \includegraphics[width=\textwidth]{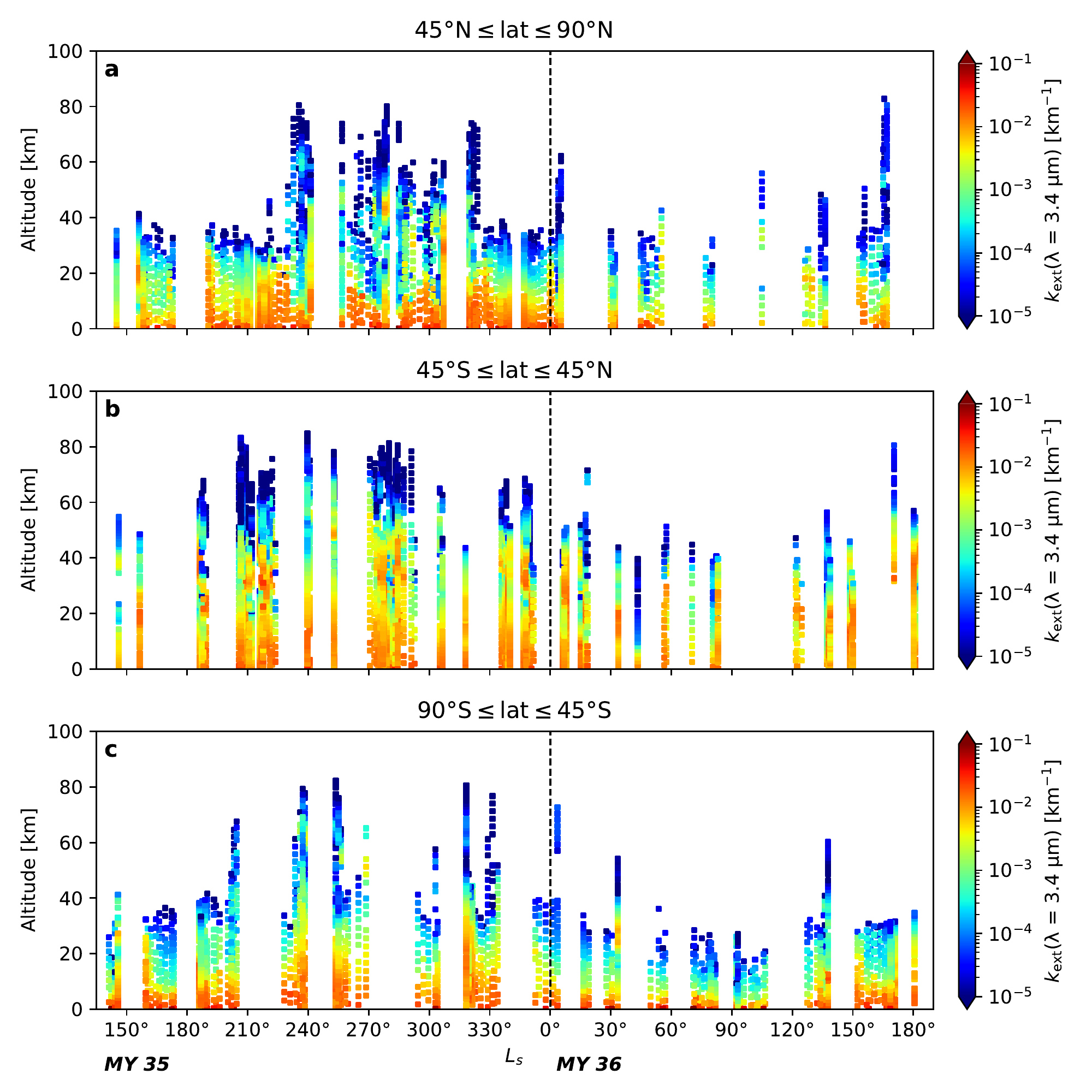}
        \caption{Vertical profiles of the measured extinction of the Martian atmosphere
          ($k_\mathrm{ext}=\mathrm{d}\tau/\mathrm{d}z$, in km$^{-1}$) at 3.4~\mum\ as observed 
          by ACS-MIR over mid-MY 35 to mid-MY 36.
          To highlight the seasonal variations of the clouds by getting rid of their latitudinal
          dependency, each panel represents the profiles grouped by ranges of latitude
          (North/Equatorial/South).}
        \label{fig:profils_kext_my35-36_filtrage_lat}
    \end{figure}

        Figures~\ref{fig:profils_kext_my35-36_filtrage_ls} \& \ref{fig:profils_kext_my35-36_filtrage_lat}
        present vertical profiles of the extinction coefficient $k_\mathrm{ext}$ at 3.4~\mum\ acquired
        by ACS-MIR between $L_s=140^\circ$ (MY 35) and $L_s=90^\circ$ (MY 36), filtered
        by latitude and $L_s$ to highlight the seasonal and latitudinal variations of the atmospheric
        vertical structure (similarly to Figures~\ref{fig:profils_reff_my35-36_filtrage_lat} \&
        \ref{fig:profils_reff_my35-36_filtrage_ls}).
        We observe in Figure~\ref{fig:profils_kext_my35-36_filtrage_ls}, similarly to what has been 
        noted above for the global behavior of the clouds, 
        that there is a strong latitudinal dependence in the atmospheric extinction profiles:
        a given $k_\mathrm{ext}$ value is observed at higher altitude close to the equator than
        in the polar regions. This is in agreement with previous retrievals of the atmospheric
        dust extinction from Mars Climate Sounder observations \cite{kleinbohl_2015}.
        This variation is only about 10~km for the lower layers of the
        atmosphere, but between $\sim30^\circ$S and $55^\circ$N we observe in most of our profiles 
        the presence of a second detached atmospheric layer (i.e., atmospheric extinction
        decreases with altitude then increases again by at least one order of magnitude)
        around $\sim$30--55~km (as previously
        noted in individual profiles from Figure~\ref{fig:multi_kext_profiles}). Thus, extinction
        values of $\sim10^{-2}$~km$^{-1}$ can be observed up to 50~km (i.e., 40~km above the altitude
        where they are observed around $60^\circ$N/S) in these layers.
        This suggests that when a large quantity of aerosols are raised above a certain altitude in the 
        atmosphere ($\sim30$~km) they do not remain in one single atmospheric layer but are split in
        two distinct layers, with a significant decrease of the atmospheric extinction (by one order
        of magnitude) between them.
        
    \subsection{Seasonal variations}
        Along with the latitudinal variations, we also present in Figure~\ref{fig:profils_kext_my35-36_filtrage_lat}
        the evolution of the atmospheric extinction profiles with the $L_s$.
        We observe that the variations of altitude for a given extinction value are of lower amplitude
        than the variations of the average water ice clouds altitude for the same profiles (cf.
        section~\ref{sec:seasonal_variations}). Between $90^\circ$S and $55^\circ$S (panel~c),
        the altitude where $k_\mathrm{ext}\sim5\cdot10^{-3}$~km$^{-1}$ goes from 5~km at 
        $L_s=180^\circ$ to 30~km at $L_s=320^\circ$, and then back to 5~km at $L_s=30^\circ$.
        We also note that when the haze top altitude increases, for a fixed range of latitudes,
        it is largely due to the expansion of low-extinction layers to the higher altitudes rather than a 
        shift of the vertical structure of the extinction profile, 
        i.e., a given extinction value remains approximately associated with the same altitude,
        but layers with lower $k_\mathrm{ext}$ values appear at the top of the profiles, 
        which is coherent with NOMAD/IUVS measurements over MY 35 \cite{streeter_2022}
        and previous SPICAM observations from MY 27 to MY 31 \cite{maattanen_2013}.
        
    \begin{figure}[htp]
        \centering
        \includegraphics[width=\textwidth]{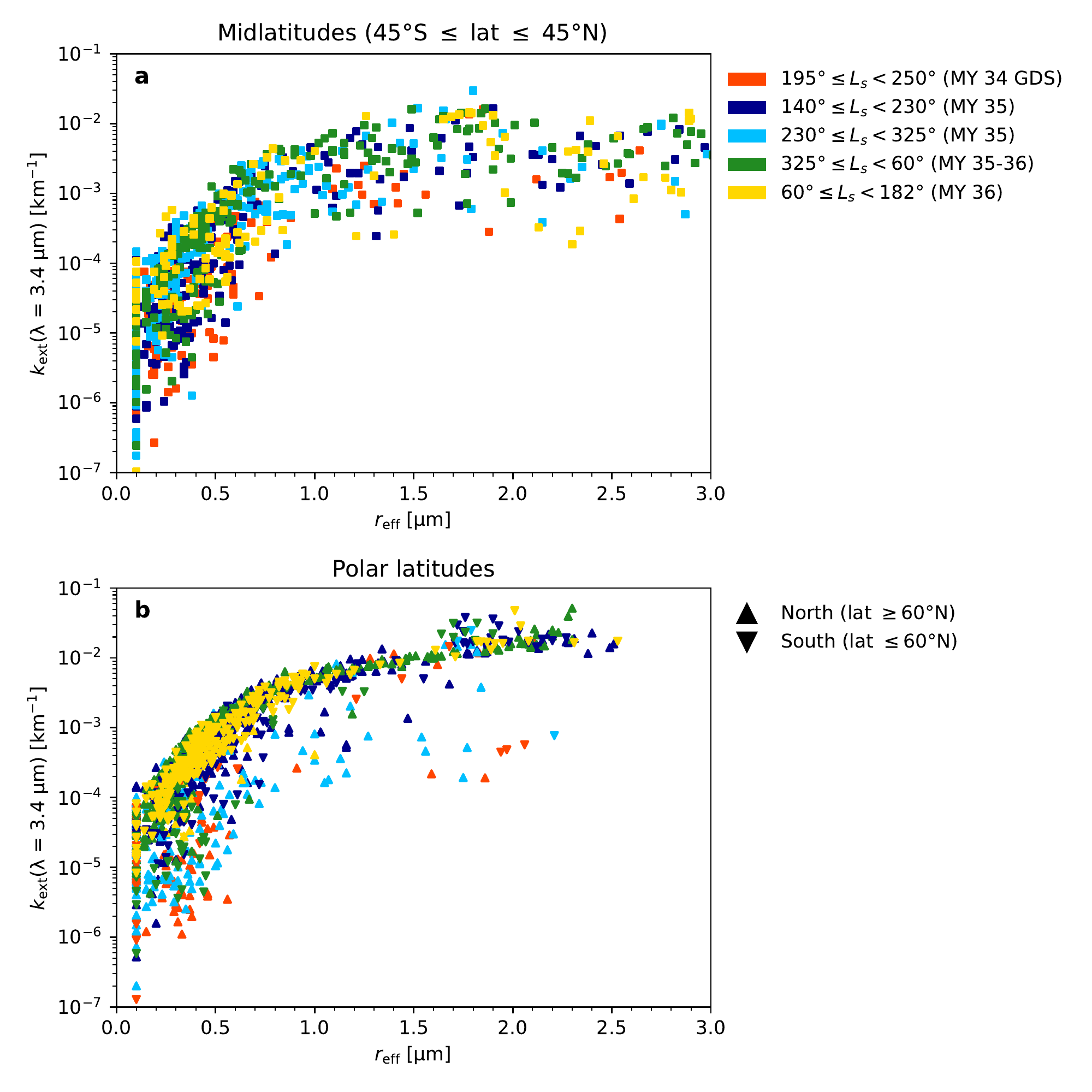}
        \caption{Distribution and size of the water ice crystals within the clouds as a function
          of their extinction coefficient $k_\mathrm{ext}$ and $L_s$ of observation, 
          for 3 ranges of latitude: midlatidudes (panel a), Northern and Southern polar latitudes
          (panel b).
          Observations acquired during the MY 34 GDS are shown in red for comparison with data
          from non-GDS years (MY 35 \& 36).}
        \label{fig:kext_reff}
    \end{figure}

    \subsection{Extinction vs crystal size}
        Figure~\ref{fig:kext_reff} presents the extinction coefficient $k_\mathrm{ext}$ of the clouds
        at $\lambda=3.4$~\mum\ as a function of the retrieved size of the water ice crystals $r_\mathrm{eff}$
        for 3 ranges of latitudes (midlatitudes, northern and southern polar latitudes) 
        and 5 ranges of $L_s$, including the MY 34 GDS.
        We observe at all latitudes the presence of 2 regimes depending on the size of the water ice
        crystals: for $r_\mathrm{eff}\leq1$~\mum\ the extinction $k_\mathrm{ext}$ increases with
        $r_\mathrm{eff}$, while for larger particles the extinction of the clouds remains in the same
        range of values when $r_\mathrm{eff}$ varies between 1.5 and 3~\mum.
        
        We can still note that, if we exclude the data from the MY 34 GDS and the MY 35 dust storm 
        (discussed further in the next paragraph), the points in the polar regions (panel b) have less scatter in terms
        of $k_\mathrm{ext}$ than in the midlatitudes (panel a), especially for the large
        particles ($r_\mathrm{eff}\geq1.5$~\mum). This may be related to the smaller variations
        of the cloud altitudes in these regions (cf. Figure~\ref{fig:alt_reff_my35-36_filtrage_ls}),
        which suggests a more stable vertical distribution of the properties of the clouds
        (both $k_\mathrm{ext}$ and $r_\mathrm{eff}$).
        Also, for fixed values of $r_\mathrm{eff}$ and $\lambda$ the extinction coefficient 
        $k_\mathrm{ext}$ is solely a function of the number of particles that have scattered the light along
        the line of sight. Thus, the small dispersion in the cloud extinction values in the polar regions
        indicates that the number density of the water ice crystals within the clouds is overall constant over
        the year.
        
        However, we observe in Figure~\ref{fig:kext_reff}b that some detections of particles with $r_\mathrm{eff}$
        between 0.9 and 2.2~\mum\ are associated with lower extinction values than the ones from the
        typical trend: $k_\mathrm{ext}\sim10^{-4}$--$10^{-3}~\mathrm{km}^{-1}$ instead of
        $k_\mathrm{ext}\sim10^{-2}~\mathrm{km}^{-1}$.
        A noteworthy point is that most of these detections correspond to observations that have been
        mostly acquired either during the MY 34 GDS or during the regional dust storm in the Northern
        hemisphere in MY 35. In addition, even for lower size of ice crystals, the GDS detections
        are correlated with lower $k_\mathrm{ext}$ values. These observations with lower
        extinction values are associated with detections at higher altitudes than those typical
        for these sizes of crystals (cf. Figure~\ref{fig:alt_reff_my35-36_filtrage_ls}b).
        Thus, while water ice clouds are observed at higher altitudes during dust storms, their
        ice crystal number density is lower than what is typically observed outside these events.

\section{Comparison with the Mars PCM}    \label{sec:comparison_models}
    Our dataset provides precise and systematic monitoring of the cloud
    altitude, effective radius, and extinction over two Martian years.
    Thus, these new observational constraints can be compared to
    predictions by numerical simulations such as GCMs. Here we will compare our results with the Mars PCM
    \cite{forget_1999, montmessin_2002a, madeleine_2012a, navarro_2014a}.
    We will focus on the observations acquired during the second half of MY 35, a non-GDS year
    for which we have access to a climatology of the atmospheric dust
    distribution \cite{montabone_2015, montabone_2020}.
    
    Even though the size of the ice crystals within the clouds is among the quantities available from version 5
    of the Mars PCM, the model sizes are systematically and significantly larger than
    the values retrieved in this study as well as in previous ones 
    \cite<e.g.,>{wolff_2003, vincendon_2011, clancy_2019, stcherbinine_2020a, luginin_2020, liuzzi_2020}.
    Investigations on this discrepancy are currently ongoing by the Mars PCM team.
    Thus, we do not discuss further the crystals' size here, but we instead focus on the altitude of the
    clouds, through the volume mixing ratio of H\textsubscript{2}O ice crystals in the atmosphere.
    
    \subsection{Global comparison for MY 35}
        \begin{figure}[ht]
            \centering
            \includegraphics[width=\textwidth]{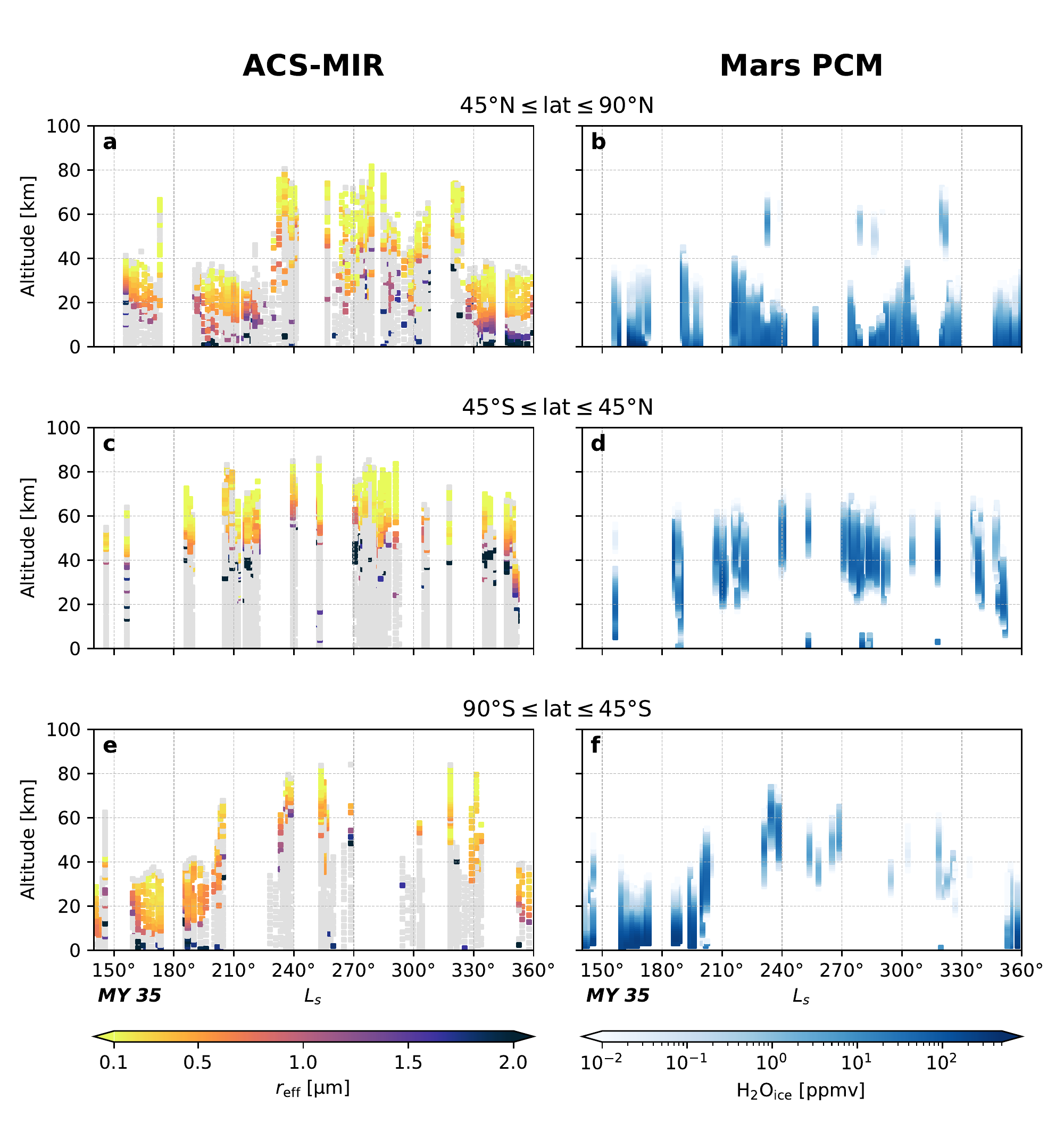}
            \caption{Comparison between the ACS-MIR cloud detections (left) and the water ice clouds
                predicted by the Mars PCM (right) over the second half of MY 35.
                Similarly to Figure~\ref{fig:profils_kext_my35-36_filtrage_lat} the profiles are
                filtered as a function of latitude in each panel to decorrelate between the
                seasonal and the latitudinal trends.}
            \label{fig:comp_acs_pcm_my35}
        \end{figure}
    
        Figure~\ref{fig:comp_acs_pcm_my35} shows the comparison between the water ice clouds detected
        by ACS-MIR (left column) and predicted by the Mars PCM (right column) from $L_s=140^\circ$
        to $L_s=360^\circ$ (MY 35) for three ranges of latitude (North and South high latitudes, mid-latitudes)
        in order to delineate the latitudinal from the seasonal trends.

        We observe that the seasonal trend shown in the ACS-MIR data for the mid-southern latitudes
        (i.e., increase of the altitude of the clouds around the perihelion) in the panels c \& e is
        well reproduced by the PCM (panels d \& f). However, the strong increase observed in the altitude of the clouds
        associated with the regional dust storm in the Northern hemisphere is only reproduced for a few model profiles (panel b).

        We also observe that in the polar regions, clouds are present in the PCM down to the surface while
        they are not detected at low altitudes in the ACS-MIR retrievals, which is not inconsistent. Indeed, as mentioned
        in \citeA{stcherbinine_2020a} our method is not sensitive to the larger ($r_\mathrm{eff}\geq2-3~\upmu$m)
        ice particles, and to mixed dust-ice layers dominated by dust.
        Thus, it is likely that we do not detect some atmospheric layers with water ice crystals at the bottom of the profiles.
        
        We observe another difference between the model and the data: the maximum altitude of the clouds in the PCM is usually lower than that observed: by $\sim$10~km in the polar regions and up to $\sim$20~km in the
        mid-latitudes. This difference is this time not linked with an observational bias (we detect more clouds layers than predicted by the model). Further investigations are required to understand why the simulations do not reproduce precisely observed altitudes.
    
    \subsection{Comparison on individual profiles}
        \begin{figure}[htp]
            \centering
            \includegraphics[width=\textwidth]{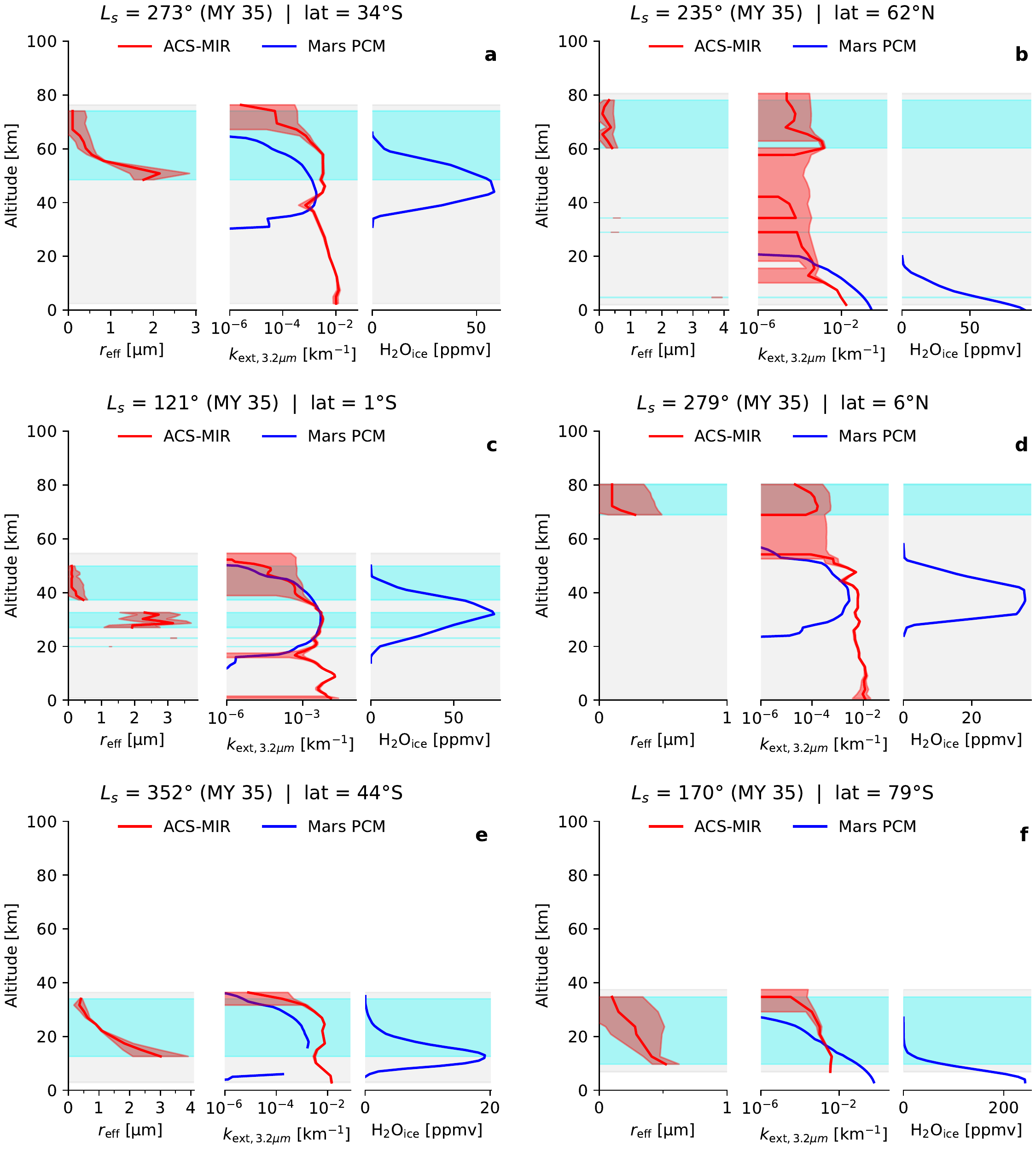}
            \caption{Comparison between the ACS-MIR retrieval (red lines) and the Mars PCM
              prediction (blue lines) for 6 individual profiles from MY 35.
              Each panel shows the water ice $r_\mathrm{eff}$ retrieved by ACS-MIR (left),
              the atmospheric extinction ($k_\mathrm{ext}$) at $\lambda=3.2~\upmu$m (center) 
              and the amount of water ice predicted by the Mars PCM (right).
              The light blue areas represent the altitudes where water ice clouds have been 
              identified in the ACS-MIR data.}
            \label{fig:comp_profiles_reff_kext_GCM}
        \end{figure}

        Figure~\ref{fig:comp_profiles_reff_kext_GCM} presents a close-up view of the comparison
        between the ACS-MIR retrievals and the Mars PCM predictions for 6 individual profiles
        taken between $L_s=121^\circ$ and $L_s=352^\circ$ in MY 35. For each profile, the information
        about the altitude of the water ice cloud in both the ACS-MIR observation and the Mars PCM
        simulations are provided, as well as the extinction coefficient at $\lambda=3.2~\upmu$m.

        We observe that the modeled extinction values ($k_\mathrm{ext}$) at $\lambda=3.2~\upmu$m predicted by the model
        match the values retrieved by ACS-MIR, at least for the altitudes
        where the PCM produces clouds (panels a, c, d \& e).
        However, we still note a shift of about 10 to 20~km in the altitudes where a cloud layer
        is typically detected versus where they are predicted by the PCM (panels
        a \& f). 
        Despite the limitations of our method, it is clear that ACS observes clouds higher than
        predicted by the PCM.
        Indeed, by looking at the ACS-MIR $k_\mathrm{ext}$ profiles, we suspect that some clouds
        layers are missing at the bottom of the clouds (panels a, e \& f) as the detached layer
        that is visible in the extinction profile extends to a lower altitude than the water ice
        cloud detection.
        However, we note that when a cloud layer is not predicted to extend as high as
        it is observed by ACS, the model atmospheric extinction also decreases faster than the
        measured one (panels a \& f), which strengthens the validity of the differences noted between
        PCM predictions and ACS observations for these altitudes.
        Plus, we also find that the largest discrepancies are observed for the 
        higher altitude clouds, and seem likely to be
        to be related to the difficulty of the model to reproduce in general the high-altitude detached layers
        of water ice detected by ACS (panels b \& d).
        
        Thus, clouds predicted at lower altitudes in the Mars PCM compared to the ACS-MIR observations
        may be the manifestation of the limits of our retrieval method, while the lack of the upper layers
        in the PCM is a shortcoming in the current model.

\section{Conclusion}    \label{sec:conclusion}
    In this paper, we present the results of our study on the identification and characterization of
    the Martian water ice clouds properties from IR SO data acquired by the ACS-MIR instrument onboard
    TGO between $L_s=163^\circ$ (MY 34) and $L_s=181^\circ$ (MY 36).
    Using the methodology previously described in \citeA{stcherbinine_2020a}, we are able to 
    simultaneously detect the presence of water ice clouds in the Martian atmosphere, and constrain
    the size of their crystals at each observed altitude.
    As TGO's orbit allows ACS to span all the latitudes between $80^\circ$S and $80^\circ$N in about
    $\sim20^\circ$ of $L_s$, this new dataset allows us to monitor the properties of the water
    ice clouds as a function of both the season and the latitude. In addition, the data acquired
    during MY 35 give us reference measurements at the same period to be compared to the observations
    acquired during the MY 34 GDS to better constrain the effects of such an event.
    
    The main results are summarized below:
    \begin{itemize}
        \item The Solar Occultation technique used by ACS-MIR provides highly sensitive measurements
            that allow us to observe optically thin clouds 
            ($k_\mathrm{ext}(\lambda=3.2~\upmu\mathrm{m}) \sim 10^{-4}~\mathrm{km}^{-1}$).
            Such clouds are typically much harder to detect through other observing geometries 
            (on-disk or limb scattering).
        \item Where thick water ice clouds appear locally in the Martian atmosphere
            \cite{smith_2013, wolff_2019, szantai_2021}, our SO observations reveal the
            quasi-systematic presence of small water ice crystals within the upper layers of
            the atmospheric aerosols (at least near the morning and evening terminators; 
            i.e., at local times $\sim$~06:00 and 18:00).
        \item The decrease of the ice crystals size as the altitude increases, previously
            noticed in the GDS study \cite{stcherbinine_2020a} and by other TGO studies
            \cite{luginin_2020, liuzzi_2020}, remains the observed behavior in this extended 2-MY dataset.
        \item The comparison between the perihelion periods of MY 34 and MY 35 allows us to
            contrast conditions during the MY 34 GDS with those of the same season in the absence 
            of such a dramatic atmospheric event. It can be seen that the global altitude of 
            the water ice clouds has increased by 10~km during the GDS compared to a more typical
            year, and this altitude increase can been 20~km for the smallest ice crystals
            ($r_\mathrm{eff}\leq0.1$~\mum).
        \item No large-grained clouds ($r_\mathrm{eff}\geq1.5$~\mum) have been observed at high altitudes (60 to 65~km range) during the non-GDS year, while they were detected during the MY 34 GDS. However, such clouds have been detected at 55~km (a relatively high altitude for such large ice crystals) close to the equator
            without being associated with a particular dust storm event.
        \item We also note that this elevation of the water ice clouds during a dust storm
            corresponds to a decrease of the typical extinction values associated with a specific
           ice crystal sizes.
        \item We observe variations of 20 to 40~km between the average altitude of the water
            ice clouds during summer and winter for both hemispheres, with water ice clouds often
            detected up to 80~km during summer seasons. The variations are stronger in the high latitudes
            compared to those in the equatorial regions.
        \item In addition to the seasonal variations, clouds are also detected typically 20
            to 40~km higher close to the equator than in the polar regions.
        \item Finally, comparison between the ACS-MIR retrievals and the Mars PCM shows
            that the water ice clouds are usually predicted at lower altitudes in the model,
            up to 10~km lower in the polar regions and 20~km in the midlatitudes.
    \end{itemize}
    
    In conclusion, we investigated the impact of season and latitude on the altitude and the size of the water ice crystals.
    These results have been compared to predictions from climate models, showing a
    good agreement overall, while revealing some regions of discrepancies associated with the cloud altitudes.
    The quasi-systematic detection of water ice clouds in our observations highlights the
    importance of considering these clouds in studies of the current Martian atmosphere and climate.

\section*{Data Availability}
Raw ACS data are available on the ESA PSA at 
\url{https://archives.esac.esa.int/psa/#!Table%20View/ACS=instrument}.
Derived particles sizes and atmospheric extinction profiles can be found in \citeA{stcherbinine_2022_dataset}.

\acknowledgments
ExoMars is a space mission of ESA and Roscosmos.
The ACS experiment is led by IKI Space Research Institute in Moscow.
The project acknowledges funding by Roscosmos and CNES.
Science operations of ACS are funded by Roscosmos and ESA.
Science support in IKI is funded by Federal agency of science organization (FANO).
MJW acknowledges support from Jet Propulsion Laboratory subcontract contract 1551112.

\bibliography{biblio_stcherbinine2022}

\end{document}